# Native NIR-emitting single colour centres in CVD diamond

**D Gatto Monticone**[1,2,3], **P Traina**[4], **E Moreva**[4], **J Forneris**[1,2,3], **P Olivero**[1,2,3]*,

**I P Degiovanni**[4], **F Taccetti**[5,6], **L Giuntini**[5,6], **G Brida**[4], **G Amato**[4] **and M Genovese**[4]

[1] Physics Department and NIS Centre of Excellence - University of Torino, Torino, Italy

[2] Istituto Nazionale di Fisica Nucleare (INFN) Sez. Torino, Torino, Italy

[3] Consorzio Nazionale Interuniversitario per le Scienze Fisiche della Materia (CNISM) - sez. Torino, Italy

[4] Istituto Nazionale di Ricerca Metrologica (INRiM), Torino, Italy

[5] Physics Department - University of Firenze, Firenze, Italy

[6] Istituto Nazionale di Fisica Nucleare (INFN) Sez. Firenze, Firenze, Italy

* corresponding author: paolo.olivero@unito.it

**Abstract.** Single-photon sources are a fundamental element for developing quantum technologies, and sources based on colour centres in diamonds are among the most promising candidates. The well-known NV centres are characterized by several limitations, thus few other defects have recently been considered. In the present work, we characterize in detail native efficient single colour centres emitting in the near infra-red ($\lambda = 740-780$ nm) in both standard IIa single-crystal and electronic-grade polycrystalline commercial CVD diamond samples. In the former case, a high-temperature (T > 1000 °C) annealing process in vacuum is necessary to induce the formation/activation of luminescent centres with good emission properties, while in the latter case the annealing process has marginal beneficial effects on the number and performances of native centres in commercially available samples. Although displaying significant variability in several photo-physical properties (emission wavelength, emission rate instabilities, saturation behaviours), these centres generally display appealing photophysical properties for applications as single photon sources: short lifetimes (0.7–3 ns), high emission rates (~50–500×$10^3$ photons $s^{-1}$) and strongly (> 95%) polarized light. The native centres are tentatively attributed to impurities incorporated in the diamond crystal during the CVD growth of high-quality type IIa samples, and offer promising perspectives in diamond-based photonics.

# 1. Introduction

Single photon sources represent a key element for developing quantum technologies [1-4]. Diamond offers a promising platform for the implementation of single-photon-emitter architectures, due to the vast range of available luminescent defects [5, 6] with suitable emission properties that can be allocated in a broadly transparent crystal structure. The nitrogen-vacancy ($NV^-$) complex established a prominent role as a single photon emitter in several pioneering works [7-9], due to its ubiquity, quantum efficiency and well-understood electronic transition structure [10]. In the last decade, single $NV^-$ emitters were successfully employed to implement quantum cryptography schemes [11-14] as well as more fundamental demonstrations of quantum complementarity and entanglement [15-18]. At the same time, the research in diamond-based single-photon sources broadened to new types of defects, with the goal of overcoming some inherent limitations in the $NV^-$ centre, namely its strong phonon coupling, relatively long lifetime and charge-state blinking. In particular, the identification of centres emitting in the near-infrared (NIR) offers the perspective of combining diamond colour centres with Si-based photodetectors in the spectral range where they are maximally efficient. On the other hand, the accessibility of centres emitting in the telecom wavelength range would offer the advantage of integrating diamond emitters with standard photonics, although it must be noted that the existence of efficient colour centres in diamond emitting at $\lambda > 1$ µm is questionable [19]. A range of different systems emitting in a more convenient NIR spectral window has been explored with regards to their fabrication strategies, physical properties and functional performances [20-23] and suitability of several centres for the implementation of quantum communication schemes has been assessed [24, 25].

In particular, the silicon-vacancy (SiV) complex has been investigated both in single-crystal [26, 27] and isolated-nanocrystal [28-32] diamond as a viable alternative to the $NV^-$ centre, due to its narrower spectral width combined with lower phonon coupling and full polarization.

With similar motivations (narrow emission, low background, shorter lifetimes, linearly-polarized NIR emission), also Ni-related (sometimes referred as NE8) and Cr-related centres have been recently investigated. The properties of Ni-related centres were studied in single-crystal [33-37], microcrystalline [38], nanocrystalline [35] and isolated-nanocrystal [39-42] diamond. Cr-related centres were characterized in single-crystal [43-46], isolated-microcrystal [47] and isolated-nanocrystal [48-50] diamond, as well. In addition to the above-mentioned centres, Xe-related [51] and H3 [52] centres have been investigated in single-crystal and isolated-nanocrystal diamond, with the perspective of further extending the range of single centres in diamond.

Recently, the activation of NIR centres upon high-temperature (T > 1000 °C) thermal annealing in forming gas (4% $H_2$ in Ar) in pristine (i.e. unimplanted) single-crystal diamonds was investigated, both in terms of creation yield and spectral emission properties [53].

It is worth noting that, differently from what observed in the $NV^-$ case, most of the above-mentioned centres frequently display zero-phonon emission lines which are dispersed on a relatively large spectral range (i.e. ~740–780 nm). This indicates that different structural configurations are possible for the same impurity and/or that the centres are extremely sensitive to local strain fields. Therefore, an unequivocal attribution of their spectral features to specific defect complexes requires particular care: to this scope, complementary techniques such as vibrionic [54] and stress-dependent [55] photoluminescence (PL) spectrometry measurements, EPR [56-58] and XANES [59] can provide useful information. As a matter of fact, several works reported promising NIR-emitting single colour centres in diamond with tentative (if any) attributions [60-62, 53].

In the present work we report on the observation and optical characterization in commercial CVD diamond samples of native NIR-emitting ($\lambda$ = 740–780 nm) single colour centres with promising photo-physical properties (linewidth, emission rate, photo-stability and polarization). While being characterized by similar properties as those reported in previous works (i.e.: high brightness, narrow emission in the $\lambda$ = 740–780 nm spectral range, ~ns radiative lifetimes, linear polarization), the centres were not formed upon ion implantation or intentional doping during CVD growth but directly observed in as-grown polycrystalline samples characterized by the highest degree of purity, while a high-temperature annealing procedure was necessary to activate/form the centres in IIa single-crystal samples characterized by a lower degree of purity. Despite a still unclear attribution, the availability of native NIR centres with good emission properties in commercial CVD samples offers numerous opportunities in diamond-based photonics.

**2. Experimental setup**

In the present study five 3×3×0.3 mm$^3$ single-crystal CVD diamond samples produced by Element Six in different lots were employed. The samples are classified as type IIa, their nominal substitutional nitrogen and boron concentrations being <1 ppm and <0.05 ppm, respectively. The crystal orientation is <100> and they are optically polished on both of their two larger faces. A 5×5×0.2 mm$^3$ polycrystalline CVD diamond sample produced by Element Six was also investigated. The sample is classified as type IIa and is referred as "detector grade" by the producer, with nominal substitutional nitrogen and boron concentrations of <50 ppb and <1 ppb, respectively. The average microcrystal size was in the 20–50 μm range.

All of the above-mentioned samples were thermally annealed in vacuum (p < 10$^{-6}$ mbar) at a temperature of 1450 °C for 1 hour. After the high-temperature annealing, surface graphitization was removed by oxidizing the sample in air for 30 min at a temperature of 400 °C, followed by a 30 min exposure to an oxygen plasma (30 W radiofrequency power, 20 sccm oxygen flux, p = 2.5×10$^{-2}$ mbar).

A single-photon-sensitive confocal NIR-PL microscopy system was developed in the present work, as schematically shown in figure 1. The centres are excited with both continuous and pulsed laser light at λ = 690 nm emitted by a solid-state laser source, coupled into a single-mode optical fibre and expanded by a 4× objective. The pulsed light excitation is performed at 80 MHz repetition rate with 100 ps pulse width; subsequently, the beam is directed to a dichroic mirror (λ > 700 nm) which reflects it to an objective (100×, NA = 0.9, air; or alternatively 100×O2, NA = 1.3, oil) focusing on the sample. The sample is mounted on a remotely controlled three-axis piezo-electric stage, allowing positioning in a 100×100 μm$^2$ area range with micrometric accuracy. The induced luminescence beam is collected together with the scattered excitation beam by the same focusing objective and then filtered through the dichroic mirror and a subsequent filters set ($\lambda > 730$ nm), thus allowing a suitable ($> 10^{12}$) attenuation of the $\lambda = 690$ nm excitation component of the beam. The filtered beam is then focused with an achromatic doublet and coupled into a graded-index multimode optical fibre, acting not only as an optical connection to the detectors, but also as the pinhole aperture for the confocal system. The fibre leads to an integrated 50:50 beam-splitter (BS) whose outputs are connected to two photon-counters based on Si-single-photon-avalanche photo-diodes (SPADs) operating in Geiger mode. This experimental configuration reproduces a "Hanbury Brown and Twiss" (HBT) interferometer [7], that allows identifying the presence of a single emitter via the measurement of the second-order autocorrelation function $g^{(2)}(t)$ from the coincidence counts between the two detectors. In the coincidence circuit the difference between the arrival times of single photons at the two detectors is measured by means of a time-to-amplitude converter whose output feeds a multi-channel analyser recording a histogram of the above-mentioned time intervals. Simultaneously, the total counts detected at the two SPADs are recorded by a digital counter, allowing a measurement of the total luminescence intensity for each pixel.

The collection of PL spectra from single centres was carried with a single-grating monochromator with 1600 grooves/mm blazed at 600 nm connected to one of the above-mentioned SPADs.

A complementary setup was developed for polarization measurements both on the absorbed and emitted radiation from single centres, where the above-mentioned dichroic mirror was replaced by a 50:50 BS and the polarization of the excitation beam was selected using a combination of a half-waveplate ($\lambda = 800$ nm) and a Glan-Taylor polarizer. For the detection of polarized PL a film-polarizer was placed before the achromatic doublet. Care was taken in subtracting slight rotation of the pump light polarization induced by the 50:50 BS as it was verified that all polarisers did not introduce a measurable beam shift during their rotation. The polarization dependence of the reflection and transmission of the BS were suitably taken into account. [63]

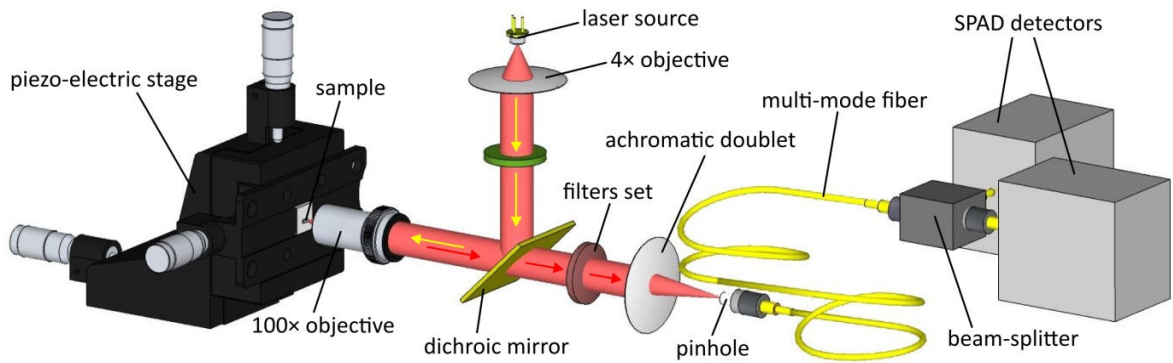

Fig. 1 (colour online): Schematic representation of the single-photon-sensitive confocal PL microscopy setup developed in the present work. The yellow arrow represents the $\lambda = 690$ nm excitation beam, while the red arrow represents the induced luminescence from the sample.

## 3. Results and discussion

*3.1. Annealing effect and spectral properties*

In single-crystal samples, thermal annealing was generally found to have the effect of inducing the formation of isolated luminescent centres in all samples under analysis.

As an example, figures 2(a) and 2(b) report confocal PL intensity maps obtained from a depth of ~5 μm below the surface in the same region of a single-crystal sample before and after 1450 °C thermal annealing, respectively. In figure 2(a), one isolated bright spot (circled in red in the map) was found to be a single emitter by means of autocorrelation statistics measurements. As shown in figure 2(b), after thermal annealing at 1450 °C, the centres concentration per surface area in single-crystal samples increased significantly, with an average value of ~1-10 centres per 50×50 μm$^2$. Due to the improvement of the emission properties (emission rate, stability) upon annealing, after thermal processing it was possible to identify and characterize a larger set of centres, with direct advantage for the statistical relevance of the acquired data. The depth distribution in the location of the centres was found to be uniform within the range of depths probed by the confocal system (~50 μm).

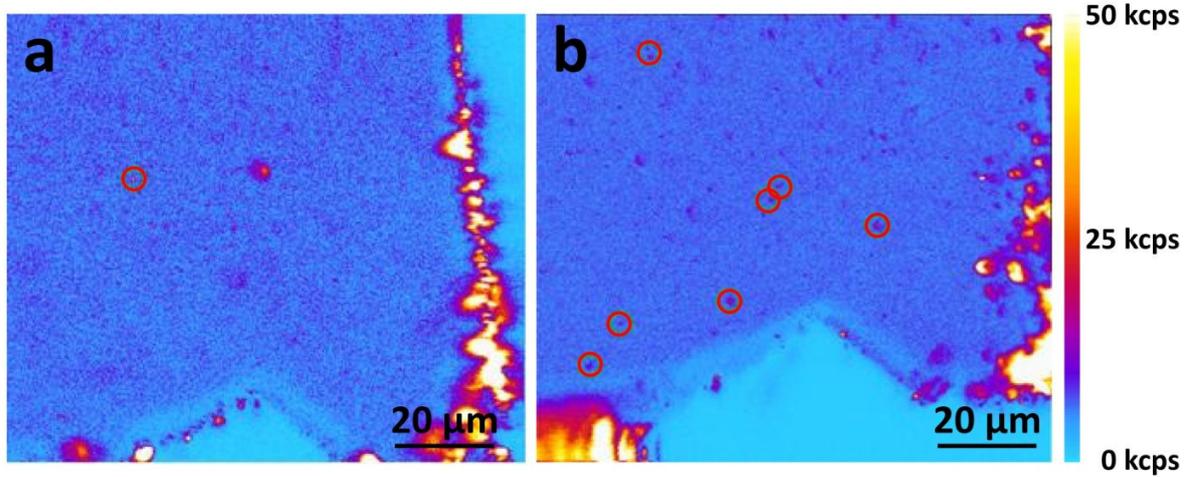

Fig. 2 (colour online): Confocal PL intensity maps acquired from one of the single-crystal samples before (a) and after (b) thermal annealing. The PL intensity is encoded in the colour scale reported on the right (0–50×10$^3$ counts s$^{-1}$). The isolated bright spots that were identified as single emitters are highlighted by the red circles.

The emission properties of the isolated centre before thermal annealing (see figure 2(a)) are reported in figure 3. Experimental data are characterized by poor statistics due to the low emission rates (~10$^4$ photons s$^{-1}$) and severe photo-bleaching of the centres. Figure 3(a) shows the PL emission of the centre at room temperature: a single zero-phonon-line (ZPL) emission at $\lambda \cong 744$ nm was observed, with two phonon sidebands. Figure 3(b) shows the relevant second-order autocorrelation chronograms, which were fitted with the following function that describes the temporal evolution of a three-level system [64]:

$$g^{(2)}(t) = 1 - (1-a) \cdot exp(-\beta_1 \cdot |t|) + a \cdot exp(-\beta_2 \cdot |t|) \qquad (1)$$

where $\beta_1$ and ($\beta_2$, $a$) account for the anti-bunching and bunching features of the chronogram, respectively. Note that the background due to diffused luminescence and Raman emission was subtracted from experimental chronograms according to the procedure described in [7] and the temporal response of the detectors was taken into account by convoluting the function in equation (1) with a Gaussian function with FWHM = 0.7 ns (as directly measured with a ps pulsed laser). A value of $(\beta_1)^{-1} = (0.77 \pm 0.07)$ ns was obtained, together with non-negligible bunching parameters: $(\beta_2)^{-1} = (0.033 \pm 0.004)$ ns and $a = (0.76 \pm 0.04)$.

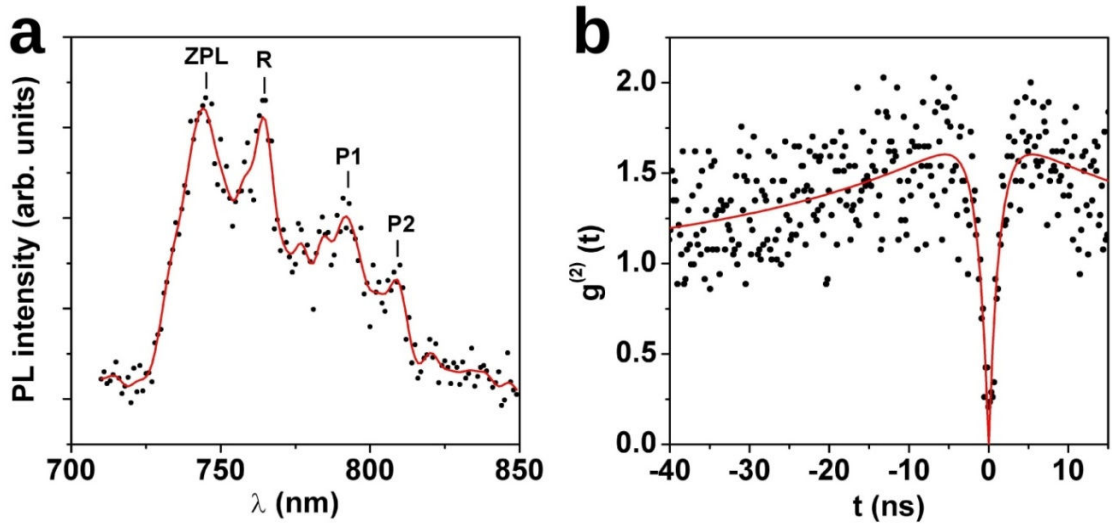

Fig. 3 (colour online): Typical PL emission properties of single colour centres in single-crystal samples before thermal annealing. a) Room-temperature PL spectrum (black dots: experimental data, red line: smoothed data for better readability): the positions of the zero-phonon line (ZPL), of the relevant phonon sidebands (P1, P2) and of the 1$^{st}$-order Raman line (R) are highlighted. b) Second-order autocorrelation chronogram (black dots: experimental data, red line: fitting function as reported in equation (1)).

The spectral position of the ZPL emission after thermal annealing (see figure 2(b)) is subjected to dispersion in the 740–780 nm range, as shown for three typical PL spectra reported in figure 4(a). No pronounced phonon sidebands were observed after thermal annealing, while partial overlap with the 1$^{st}$-order Raman line was possible, depending from the ZPL position. It is worth noting that the spectral resolution of the spectrometer poses a $\Delta\lambda = 5$ nm upper limit to the ZPL emission width at room-temperature. Due to time constraints, a limited number of centres (i.e. ~50) was characterized. Nonetheless, within the limitation imposed by statistics, from the ZPL position histogram reported in figure 4(b) it can be noted that i) the distribution of the ZPL positions is not uniform in the above-mentioned spectral range but rather multi-modal, with larger populations for $\lambda \cong 750$ nm, $\lambda \cong 755$ nm and $\lambda \cong 765$ nm; ii) the distribution does not vary significantly from sample to sample. In our interpretation, the observation of a multi-modal distribution of the ZPL spectral positions indicates that a discrete set of different physical conditions (defect structure or charge state, local strain) determines different PL emission properties.

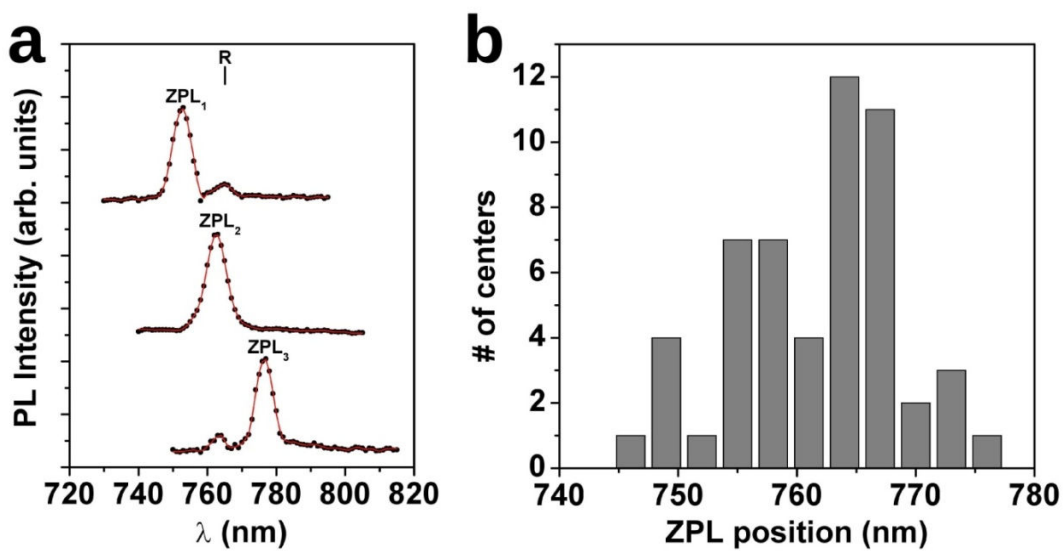

Fig. 4 (colour online): (a) Typical room-temperature PL spectra collected from different single centres in single-crystal samples after thermal annealing; the spectral positions of the ZPL peaks (ZPL$_\#$), as well as of the 1$^{st}$-order Raman line (R), are highlighted. The position of the ZPL emission exhibits variability in the 740–780 nm range, as summarized in the statistics of the ZPL spectral position (b).

With the scope of investigating the role of vacancy-related defects in the formation of the observed centres, a single-crystal sample was irradiated with 15.6 MeV C$^{5+}$ ions prior to thermal annealing. The implantation was performed at the scanning ion microbeam line of the LABEC laboratories (Florence) [65, 66] at fluences ranging from $1\times10^{13}$ cm$^{-2}$ to $1\times10^{16}$ cm$^{-2}$ over areas with sizes ranging from $50\times50$ μm$^2$ to $750\times750$ μm$^2$. No significant differences were found in the distribution and PL emission properties of single centres located in the implanted portions of the irradiated sample with respect to both the unimplanted portions and to the unirradiated samples. Therefore, it is reasonable to conclude that vacancies do not play any significant role in the formation of the colour centres under investigation: this observation is compatible with what has been reported in [53]. If the limited diffusivity of foreign atoms in diamond is considered [67, 68], the significant increase in colour centres after annealing can be hardly related to the migration of impurities over significant distances, but rather to their local rearrangement and/or to the removal of nearby PL-quenching defects. In [53], the presence of hydrogen in the atmosphere during the 1000 °C annealing was found to have a significant correlation with the formation of NIR-emitting centres with similar emission properties to those reported in the present work. Although hydrogen was never intentionally introduced in the annealing environment in the present work, the effect of trace H impurities in the process chamber cannot be ruled out in principle. Moreover, it is worth noting that in the present work the higher (1450 °C) annealing temperature could more have more effectively contributed to the dissociation of molecular hydrogen into atomic hydrogen.

In the poly-crystalline sample, native NIR-emitting single centres with good emission properties (i.e. count rates up to $1\times10^6$ photons s$^{-1}$ and absence of bunching behaviour in ~90% of observed emitters) were found even before thermal annealing (see figure 5). This observation is compatible with the higher purity of this sample, i.e. in this case thermal annealing is not necessary to remove nearby PL-quenching defects. Moreover, it is worth noting that the observation of the centres in as-grown poly-crystalline samples might be related to the effect of a higher hydrogen concentration at the grain boundaries [69]. The centres were found to be often located in the proximity of dislocations and grain boundaries. Thermal annealing at 1450 °C was found to only have a slightly positive effect of the photo-physical properties of the centres.

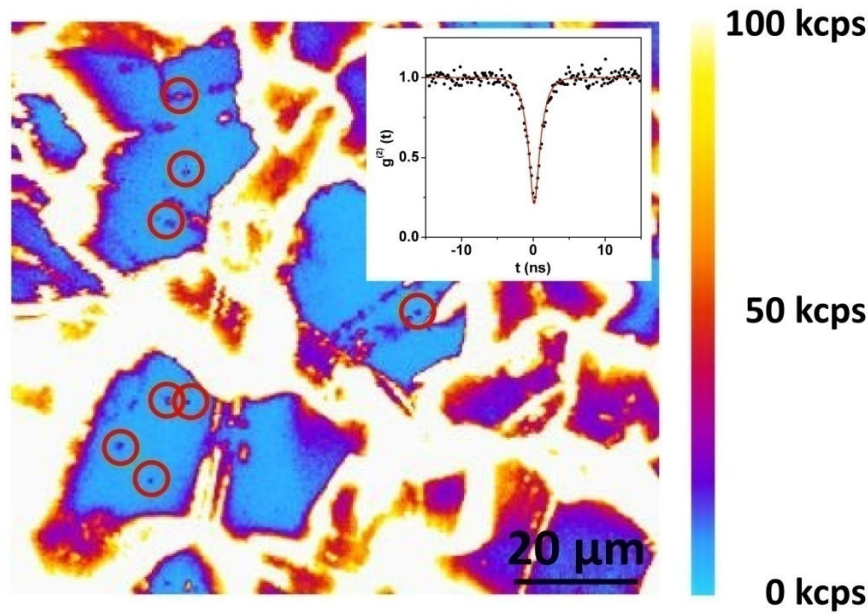

Fig. 5 (colour online): Confocal PL intensity maps acquired from the polycrystalline sample before thermal annealing. The PL intensity is encoded in the colour scale reported on the right (0–100×$10^3$ counts s$^{-1}$). The isolated bright spots, identified as single emitters, are highlighted by the red circles. A typical second-order autocorrelation function is shown in the inset.

*3.2. Emission properties*

PL instability features such as photo-bleaching and photo-blinking affect many (> 50%) of the single centres even after thermal annealing. This effect is particularly evident in standard type IIa single-crystal samples with respect to the "detector-grade" poly-crystalline sample. Similarly to what was reported for the spectral dispersion in the ZPL emission, different emission-instability features are observed from centre to centre: in some cases sudden random jumps in emission rates take place, while in others clear evidence of photo-blinking is observable.

An example of the former phenomenon is shown in figure 6(a), in which the time evolution of the emission rate of a single centre in one of the single-crystal samples is reported for a laser pump power of ~1.2 mW.

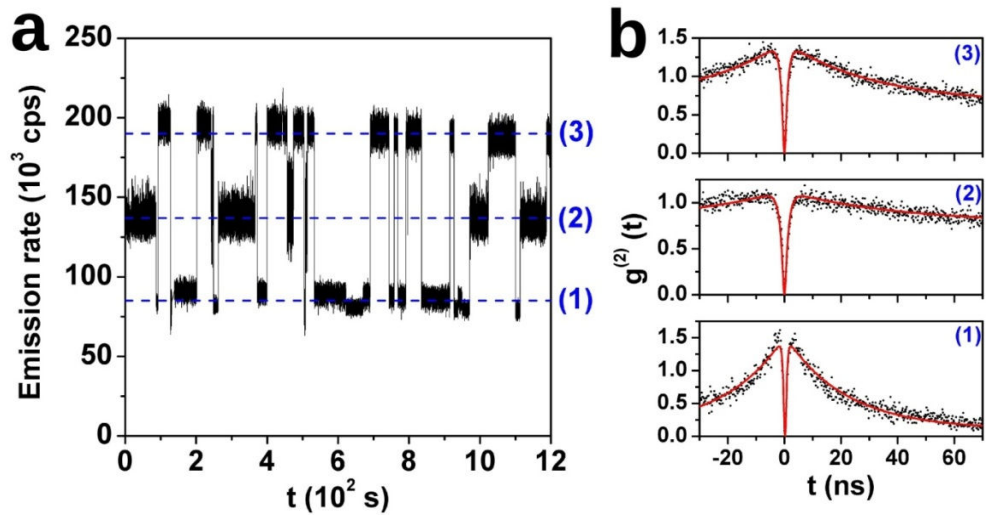

Fig. 6 (colour online): (a) temporal evolution of the emission rate from a single centre in one of the single-crystal samples after thermal annealing; laser pump power was ~1.2 mW; random oscillations among three distinct emission regimes are clearly visible, as highlighted by the three blue dashed lines labeled at the right vertical axis as (1)-(3). (b) Corresponding second-order autocorrelation chronograms measured from the same centre in each specific emission rate regime. Experimental data are reported in black dots and the corresponding fitting functions (see equation (1)) are reported in red lines.

Apart from a slight overall decrease of the count rate, which is attributed to instrumental features (i.e. a slight drift of the sample stage), the system is randomly oscillating among three distinct emission rate regimes, highlighted by the blue dashed lines and labeled as (1)-(3). This kind of process was already observed in both Ni/Si-related [37] and Cr-based [46] single centres in diamond and in the present work (as much as in [62]) is tentatively attributed to the presence of neighboring defects with unstable charge state that activate/deactivate non-radiative decay paths via resonant energy transfer, although this explanation of the observed process cannot be considered as conclusive. Figure 6(b) shows the second-order autocorrelation chronograms measured from the same centre in time windows characterized by each of the above-mentioned emission rate regimes. Experimental data were interpolated with equation (1), as for the before-annealing case. The correlation between bunching/antibunching behaviours of the autocorrelation functions and the relevant emission regimes is not straightforward, thus making their interpretation difficult: as indicated by the fitting parameters reported in table 1, i) for the lowest emission rate (labeled as (1) in figure 6(a)) a very strong bunching behaviour is observed; ii) for the intermediate emission rate (labeled as (2) in figure 6(a)) the bunching behaviour is fairly negligible and finally iii) the highest emission rate (labeled as (3) in figure 6(a)) is associated with a non-negligible bunching behaviour.

**Table 1.** Fitting parameters (see equation (1)) of the second-order autocorrelation functions reported in figure 6(b) in correspondence of different emission rate regimes (see figure 6(a)).

| Emission rate ($10^3$ cps) | $(\beta_1)^{-1}$ (ns) | $(\beta_2)^{-1}$ (ns) | a |
|---|---|---|---|
| ~85 | $2.25 \pm 0.08$ | $(4.68 \pm 0.07) \times 10^{-2}$ | $1.420 \pm 0.014$ |
| ~135 | $(8.27 \pm 0.18) \times 10^{-1}$ | $(2.34 \pm 0.07) \times 10^{-2}$ | $(3.5 \pm 0.6) \times 10^{-2}$ |
| ~190 | $1.00 \pm 0.02$ | $(3.1 \pm 0.5) \times 10^{-2}$ | $(8.02 \pm 0.08) \times 10^{-1}$ |

Although only tentatively, the above mentioned behaviour can be attributed to quantum efficiency fluctuations induced by interaction with surrounding defects [70] affected by reversible changes in their charge state [37]: these effects induce "on-off" periods of the colour centres of the order of tens of ns, as resulting from the fitting of the "bunching" components of the second-order autocorrelation functions (figure 6(b)). Moreover as a partial support of this interpretation it is worth noting that, although this kind of measurement was not possible with our experimental setup, random "jumps" among different emission regimes were previously associated to spectral shifts in ZPL emission [46, 62]. Centres subjected to the above-mentioned instabilities can bleach (i.e. "jump" into a state with extremely low emission rate) over indefinitely long times. Usually they can be retrieved by exposing them for ~10-20 s to a $\lambda = 532$ nm CW excitation with ~3 mW power.

Figure 7 reports examples of the different Poissonian/blinking emission behaviours observed under ~1 mW laser excitation from different centres in one of the single-crystal samples after thermal annealing.

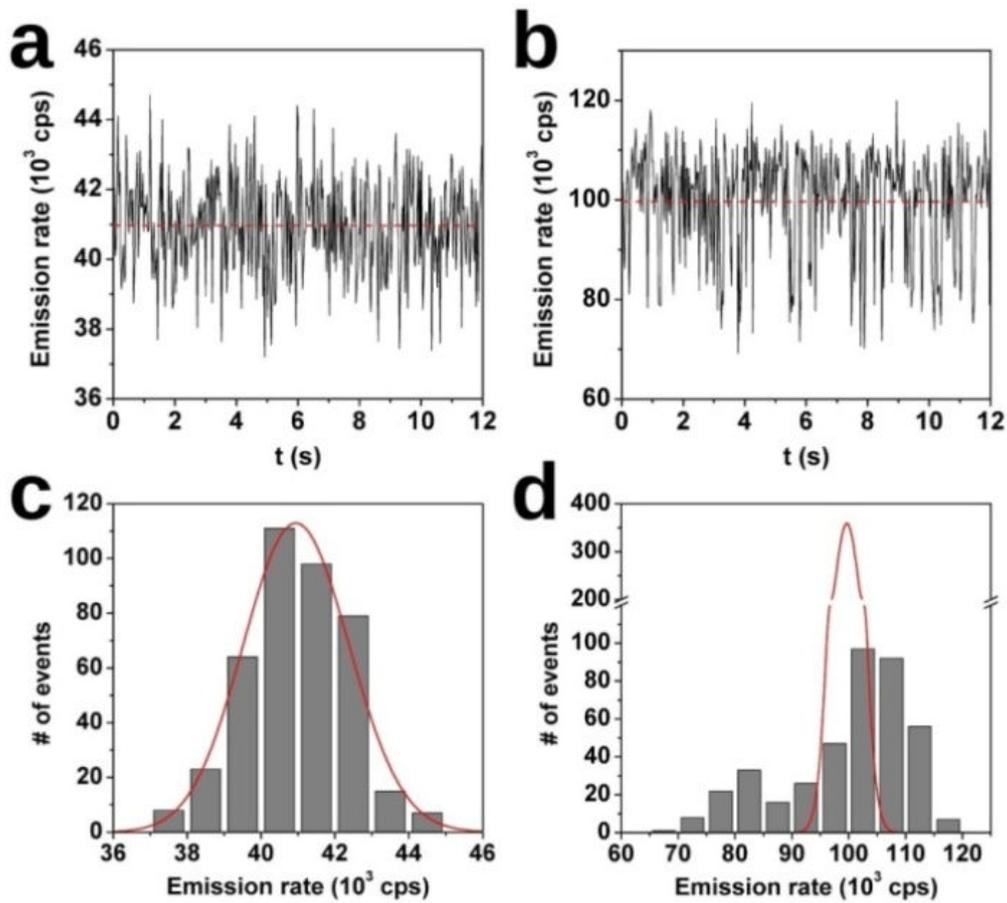

Fig. 7 (colour online): (a, c) temporal evolution of the emission rate from two single centres in one of the single-crystal samples after thermal annealing; laser pump power was ~1 mW; in both plots, the continuous black lines report the experimental data, while the dashed red lines indicate the respective mean values. (b, d) respective emission rate histograms obtained by adopting 20 ms time bins (gray vertical bars) as compared with the corresponding Poissonian distributions associated with the above-mentioned mean values. In the case of the former centre (a, b), the experimental emission rate distribution is compatible with Poissonian statistics, while in the latter case (c, d) a significant discrepancy is observed.

In figures 7(a) and 7(b), the continuous black lines report the temporal evolution of the emission rate from the two centres (the integration time for each datapoint is 20 ms), while the dashed red lines show the respective mean values. As shown in figures 7(c) and 7(d), the respective emission rate histograms (gray vertical bars) display different emission rate distributions associated with the above-mentioned mean values (continuous red lines). In particular, in the case of the former centre (figures 7(a) and 7(c)) the emission rate distribution is fully compatible with Poissonian statistics: assuming that the distortion effects on photon statistics due to SPAD dead-time (i.e. 50 ns) is negligible at the measured count rate [73], we conclude that the emission rate of the first colour centre is constant in time. On the other hand, in the latter case (figures 7(b) and 7(d)) a significant discrepancy from a Poissonian distribution is clearly observable, suggesting that the photon emission rate fluctuates in time. For comparison, a Poissonian distribution of a colour centre that would have the same intensity but with a constant photon emission rate is superimposed in figures 7(c) and 7(d). As mentioned above, a possible interpretation of this phenomenon may reside in the quenching mechanisms between the centres and surrounding defects, which determine "on-off" periods on a different time-scale with respect to what observed in figure 6 (i.e. ~$10^2$ μs).

Second-order autocorrelation measurements were carried out as a function of excitation power in centres unaffected by emission instabilities, with the purpose of estimating their excited-state lifetimes. A degree of variability was observed from centre to centre also in this respect. As shown in figures 8(a) and 8(d), centres with ZPL emission centered at different wavelengths ($\lambda_{ZPL}$ = 755 nm and $\lambda_{ZPL}$ = 772 nm, respectively) exhibit distinct saturation behaviours in the "emission rate vs pump power" trend, as resulting from the fitting of the experimental data with the following expression:

$$I(P) = I_\infty \cdot \frac{P}{P+P_{sat}} \qquad (2)$$

where *I* is the count rate, *P* is the pump power and ($I_\infty$, $P_{sat}$) are the fitting parameters. From the fitting of the data reported in figures 8(a) and 8(d), the parameters ($I_\infty = (2.5 \pm 0.4) \times 10^5$ photons s$^{-1}$, $P_{sat} = (3.7 \pm 0.7)$ mW) and ($I_\infty = (4.25 \pm 0.05) \times 10^4$ photons s$^{-1}$, $P_{sat} = (0.7 \pm 0.2)$ mW) are obtained respectively, indicating a stronger saturation behaviour in the former case.

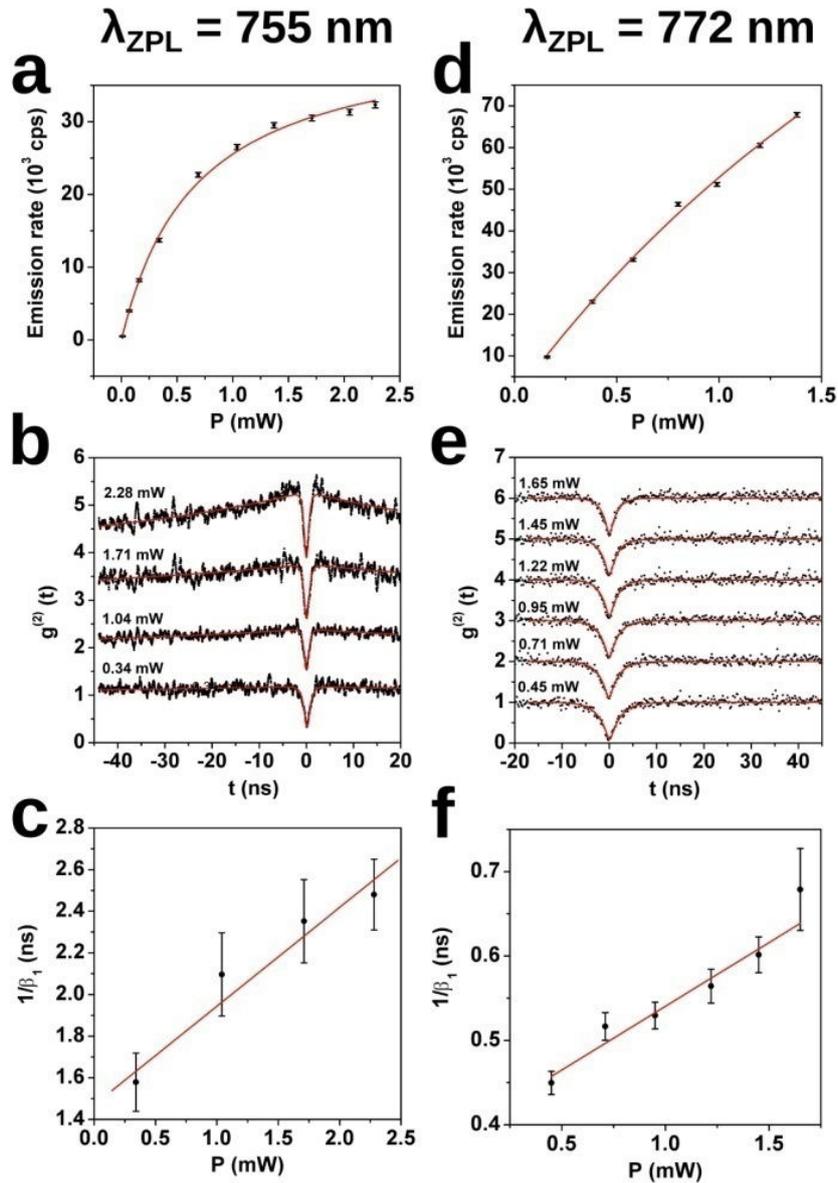

Fig. 8 (colour online): photo-physical properties of two different centres with ZPL emissions centered at $\lambda = 755$ nm (a, b, c) and $\lambda = 772$ nm (d, e, f), respectively. Experimental data are reported in black dots, while the relevant fitting curves are reported in red lines. Different saturation behaviours are observed in the emission rate vs excitation power trend (a, d), as well as in the bunching behaviour of relevant second-order autocorrelation chronograms (b, e; note: the second-order autocorrelation chronograms are displaced along the vertical axis with offsets of 1, for sake of readability) and in the corresponding lifetimes derived from the linear fitting of the "$1/\beta_1$ vs power" plots (c,f).

This indicates either that distinct non-radiative decay paths are present in different centres or that they are characterized by significantly different absorption cross sections. The second-order autocorrelation function was also measured at different powers from the same centres (figures 8(b) and 8(e)) and from the linear fitting of the "$1/\beta_1$ vs pump power" trends it was possible to estimate the centre's lifetime [60, 39, 61]. As shown in figures 8(c) and 8(f), the two centres exhibit different lifetimes, i.e. (0.7 ± 0.2) ns and (2.6 ± 0.2) ns respectively. It is worth noting that, although displaying some dispersion, all lifetime values are comprised between 0.7 ns and 3 ns, over a population of ~20 probed centres.

PL measurements were also performed under pulsed excitation, so that the lifetime was directly extrapolated by fitting the relevant chronograms. Figure 9 shows the results of second-order autocorrelation measurements carried in pulsed excitation conditions from a single centre in one of the single-crystal samples after thermal annealing.

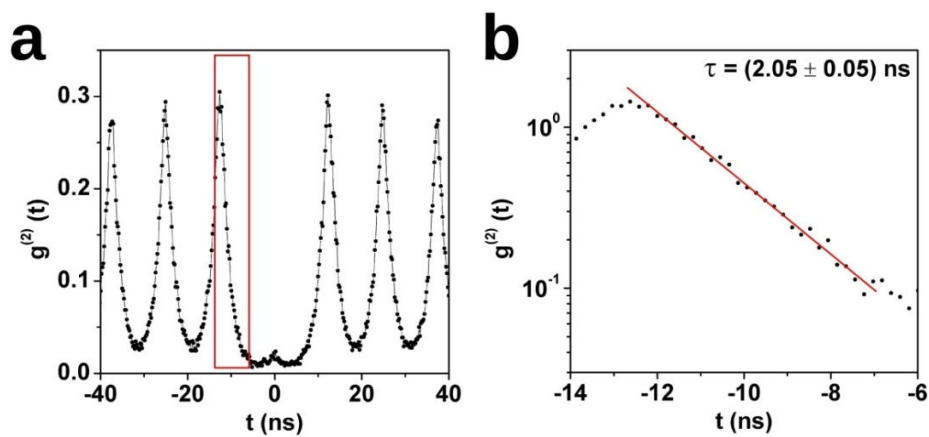

Fig. 9 (colour online): (a) second-order autocorrelation chronogram obtained from a single NIR centre in one of the single-crystal samples after thermal annealing: the spacing between the peaks of the $g^{(2)}(t)$ function correspond to the period of the pulsed excitation, i.e. 12.5 ns, while at null time delay the observed weak peak is due to a unfiltered 1st-order Raman emission. (b) zoomed-in portion of the data highlighted by the red rectangle in (a): the semi-logarithmic plot highlights the single-exponential decay behaviour; the fitting procedure (red line) yields the lifetime indicated in the inset.

As shown in figure 9(a), the spacing between the peaks of the $g^{(2)}(t)$ function corresponds to the period of the pulsed excitation, i.e. 12.5 ns, while at null time delay the observed weak peak is due to unfiltered first-order Raman emission. The zoomed-in autocorrelation chronogram reported in semi-logarithmic scale in figure 9(b) exhibits a single-exponential decay behaviour. From the fitting of the above-mentioned exponential decay, a lifetime of $(2.05 \pm 0.05)$ ns is obtained, in good agreement with typical values obtained from the above-mentioned power-dependent measurements.

*3.3. Polarization measurements*

Polarization measurements from different centres were performed both in absorption and reflection. In the former case, the full (i.e. over all polarizations) PL emission intensity of the centre in the $\lambda > 730$ nm spectral range was measured as a function of the excitation light polarization. In the latter case the PL emission intensity was measured at different polarizations while keeping the excitation polarization in the direction that maximizes the PL yield. Despite the above-mentioned variability in several photo-physical characteristics, all observed centres consistently displayed a dipole behaviour both in absorption and emission, with the polarization axis aligned with the <110> or <111> crystallographic direction. A high degree of polarization (> 95%, basically limited by the instrumental sensitivity of the setup) was observed in all centres under investigation, a significant property for quantum technology applications. In figures 10(a), (b) and 10(c), (d) the absorption and emission polarization diagrams for the two above-mentioned centres emitting respectively at $\lambda_{ZPL} = 755$ nm and $\lambda_{ZPL} = 772$ nm are reported. As shown in figure 10, in the former case ($\lambda_{ZPL} = 755$ nm: figures 10(a) and 10(b)) the excitation and emission polarizations are parallel, while in the latter ($\lambda_{ZPL} = 772$ nm: figures 10(c) and 10(d)) they are mutually orthogonal. These mutual orientations were systematically observed in all centres under investigation.

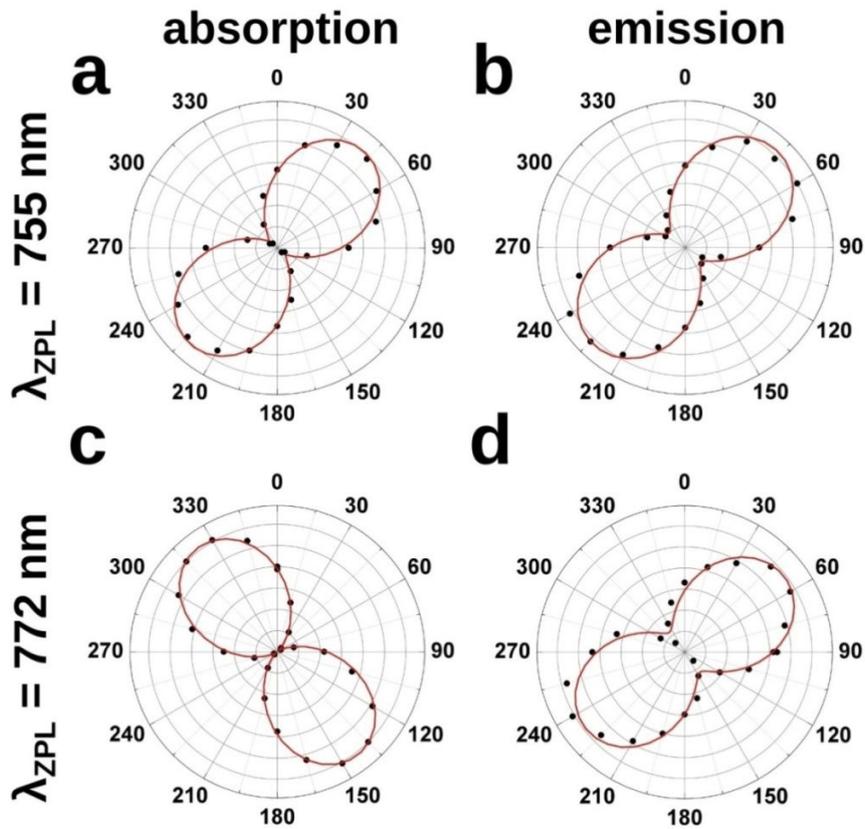

Fig. 10 (colour online): polarization diagrams from the same centres with ZPL emissions at λ = 755 nm (a, b) and λ = 772 nm (c, d) reported in figure 9. The horizontal and vertical axes (i.e. the 0° and 90° directions in the diagrams) correspond to the [100] main crystallographic axis of the sample, which was carefully oriented during the measurements. Polarization measurements were performed both on the absorbed (a, c) and emitted (b, d) light.

## 4. Conclusions

We reported about the observation and photo-physical characterization of NIR-emitting single colour centres in both optical-grade single-crystal and electronic-grade poly-crystalline commercial CVD diamond samples. In the former case, a high-temperature (T > 1000 °C) annealing process in vacuum was necessary to induce the formation/activation of luminescent centres with good emission properties, while in the latter case the annealing process had marginal effects on the number and performances of native centres in commercially available samples. While no definite hypotheses can be drawn about the attribution of the centres to specific defect complexes, our observation indicate that they are based on native impurities and/or structural defects in high-quality commercial CVD samples, with the possible exception of light elements present in trace concentrations in the annealing chamber.

The properties of the centres can be summarized as follows:

- their PL emission is characterized by narrow ZPL lines, with low background and no measurable phonon sidebands, such a property representing a fundamental advantage respect to $NV^-$ centres as single-photon emitters;

- the ZPL positions are dispersed over a relatively broad spectral range, with a multi-modal statistical distribution; this could, in principle, represent a limitation for quantum technology applications, nevertheless this issue can be addressed when the attribution of these centres is fully clarified;

- while the typical emission rate of the centres is high as compared to "standard" $NV^-$ centres [71, 72], different centres display a variety of emission instabilities; the centres unaffected by blinking behaviours display Poissonian statistics;

- also the saturation behaviour as a function of excitation power displays significant variability; also in this case, significant improvements in addressing this issue will be possible once the attribution of the centres is clarified;

- all centres are characterized by lifetime values comprised between 0.7 ns and 3 ns and their emission is linearly polarized.

It is worth noting that the observed centres have similar properties to those of previously reported emitters obtained in diamond by direct ion implantation or intentional doping during CVD growth and characterized by different tentative attributions. While a definitive defect attribution cannot be presented in the present work, the obtained results indicate that i) the structure of the observed defects is related to native impurities incorporated during high-purity (i.e. "electronic grade") CVD growth; ii) single substitutional vacancies do not have direct influence on the defect formation and iii) hydrogen might play a role in the formation of the defect. The photophysical properties of the centres indicate that more common defects (such as the NV centre or other nitrogen-related complexes) play a "quenching" role in reducing the quantum efficiency of these centres via resonant energy transfer in samples characterized by lower purity standards.

The observation in commercially available "standard type IIa" and "electronic-grade" CVD diamonds of efficient native NIR colour centres emitting fully polarized light in a convenient spectral range, with no need of specific sample processing (i.e. doping from vapour phase or by ion implantation) offers appealing perspectives in the fields of diamond-based single-photon emitters and quantum optics.

**Acknowledgments**

The authors wish to thank S. Pezzagna (University of Leipzig), P. Spinicelli (ESPCI Paris Tech Exchange) and B. Gibson (RMIT Melbourne) for the useful suggestions in the setup of the confocal microscopy setup. E. Bernardi (University of Torino) is gratefully acknowledged for his support in samples preparation. This research activity was funded by the following projects, which are gratefully acknowledged: FIRB "Future in Research 2010" project (CUP code: D11J11000450001) funded by the Italian Ministry for Teaching, University and Research (MIUR); EMRP projects "EXL02-SIQUTE" and "IND06-MIQC" (jointly funded by the EMRP participating countries within EURAMET and the European Union); "A.Di.N-Tech." project (CUP code: D15E13000130003) funded by University of Torino and Compagnia di San Paolo in the framework of the "Progetti di ricerca di Ateneo 2012" scheme; NATO SPS Project 984397; "Compagnia di San Paolo" project "Beyond classical limits in measurements by means of quantum correlations"; one-month DAAD 2011 type-1B grant "Ion implantation in diamond for applications in photonics" (application number: A/11/78148) funded by the German Academic Exchange Service.

**References**

[1] Oxborrow M and Sinclair A G 2005 *Contemp. Phys.* **46** 173−206

[2] Eisaman M D, Fan J, Migdall A and Polyakov S V 2011 *Rev. Sci. Instrum.* **82** 071101

[3] Brida G et al. 2012 *Appl. Phys. Lett.* **101 (22)** 221112

[4] Lasota M, Demkowicz-Dobrzanski R and Banaszek K 2013 *Int. J. Quantum Inf.* **11 (3)** 1350034

[5] Zaitsev A M 2001 *Optical Properties of Diamond - Data Handbook* (New York: Springer)

[6] Dischler B 2012 *Handbook of Spectral Lines in Diamond: Volume 1: Tables and Interpretations* (New York: Springer)

[7] Kurtsiefer C, Mayer S, Zarda P and Weinfurter H 2000 *Phys. Rev. Lett.* **85 (2)** 290−3

[8] Beveratos A, Brouri R, Gacoin T, Poizat J-P and Grangier P 2001 *Phys. Rev. A* **64** 061802

[9] Beveratos A, Kühn S, Brouri R, Gacoin T, Poizat J-P and Grangier P 2002 *Eur. Phys. J. D* **18** 191−6

[10] Acosta V and Hemmer P 2013 *MRS Bull.* **38** 127−30

[11] Beveratos A, Brouri R, Gacoin T, Villing A, Poizat J-P and Grangier P 2002 *Phys. Rev. Lett.* **89 (18)** 187901

[12] Alléaume R, Treussart F, Messin G, Dumeige Y, Roch J-F, Beveratos A, Brouri-Tualle R, Poizat J-P and Grangier P 2004 *New J. Phys.* **6** 92

[13] Childress L, Taylor J M, Sørensen A S and Lukin M D 2006 *Phys. Rev. Lett.* **96** 070504

[14] Schmunk W et al. 2012 *Metrologia* **49** S156−60

[15] Braig C, Zarda P, Kurtsiefer C and Weinfurter H 2003 *Appl. Phys. B* **76** 113−6

[16] Sipahigil A, Goldman M L, Togan E, Chu Y, Markham M, Twitchen D J, Zibrov A S, Kubanek A and Lukin M D 2012 *Phys. Rev. Lett.* **108** 143601

[17] Bernien H, Childress L, Robledo L, Markham M, Twitchen D and Hanson R 2012 *Phys. Rev. Lett.* **108** 043604

[18] Bernien H et al. 2013 *Nature* **497** 86−90

[19] Rogers L 2010 *Phys. Procedia* **3** 1557−61

[20] Aharonovich I and Prawer S 2010 *Diam. Relat. Mater.* **19** 729−33

[21] Orwa J O, Greentree A D, Aharonovich I, Alves A D C, Van Donkelaar J, Stacey A and Prawer S 2010 *J. Lumin.* **130** 1646−54

[22] Pezzagna S, Rogalla D, Wildanger D, Meijer J and Zaitsev A 2011 *New J. Phys.* **13** 035024

[23] Aharonovich I, Castelletto S, Simpson D A, Su C-H, Greentree A D and Prawer S 2011 *Rep. Prog. Phys.* **74** 076501

[24] Su C-H, Greentree A D and Hollenberg L C L 2009 *Phys. Rev. A* **80** 052308

[25] Neu E, Agio M and Becher C 2012 *Opt. Express* **20 (18)** 19956−71

[26] Wang C, Kurtsiefer C, Weinfurter H and Burchard B 2006 *J. Phys. B At. Mol. Opt. Phys.* **39** 37−41

[27] Rogers L J et al. 2013 *arXiv* 1310.3804v1 [quant-ph]

[28] Neu E, Steinmetz D, Riedrich-Möller J, Gsell S, Fischer M, Schreck M and Becher C 2011 *New J. Phys.* **13** 025012

[29] Neu E et al. 2011 *Appl. Phys. Lett.* **98** 243107

[30] Neu E, Fischer M, Gsell S, Schreck M and Becher C 2011 *Phys. Rev. B* **84** 205211

[31] Neu E, Albrecht R, Fischer M, Gsell S, Schreck M and Becher C 2012 *Phys. Rev. B* **85** 245207

[32] Neu E, Hepp C, Hauschild M, Gsell S, Fischer M, Sternschulte H, Steinmuller-Nethl D, Schreck M and Becher C 2013 *New J. Phys.* **15** 043005

[33] Gaebel T, Popa I, Gruber A, Domhan M, Jelezko F and Wrachtrup J 2004 *New J. Phys.* **6** 98

[34] Wu E, Jacques V, Treussart F, Zeng H, Grangier P and Roch J-F 2006 *J. Lumin.* **119–120** 19−23

[35] Wolfer M, Kriele A, Williams O A, Obloh H, Leancu C-C, Nebel C E 2009 *Phys. Status Solidi A* **206 (9)** 2012−15

[36] Orwa J O, Aharonovich I, Jelezko F, Balasubramanian G, Balog P, Markham M, Twitchen D J, Greentree A D and Prawer S 2010 *J. Appl. Phys.* **107** 093512

[37] Steinmetz D, Neu E, Meijer J, Bolse W and Becher C 2011 *Appl. Phys. B* **102** 451−8

[38] Rabeau J R, Chin Y L, Prawer S, Jelezko F, Gaebel T and Wrachtrup J 2005 *Appl. Phys. Lett.* **86** 131926

[39] Wu E, Rabeau J R, Roger G, Treussart F, Zeng H, Grangier P, Prawer S and Roch J-F 2007 *New J. Phys.* **9** 434

[40] Aharonovich I, Zhou C, Stacey A, Treussart F, Roch J-F and Prawer S 2008 *Appl. Phys. Lett.* **93** 243112

[41] Aharonovich I, Zhou C, Stacey A, Orwa J, Castelletto S, Simpson D, Greentree A D, Treussart F, Roch J-F and Prawer S 2009 *Phys. Rev. B* **79** 235316

[42] Marshall G D, Gaebel T, Matthews J C F, Enderlein J, O'Brien J L, and Rabeau J R 2011 *New J. Phys.* **13** 055016

[43] Aharonovich I, Castelletto S, Johnson B C, McCallum J C, Simpson D A, Greentree A D and Prawer S 2010 *Phys. Rev. B* **81** 121201

[44] Castelletto S, Aharonovich I, Gibson B C, Johnson B C and Prawer S 2010 *Phys. Rev. Lett.* **105** 217403

[45] Aharonovich I, Castelletto S, Johnson B C, McCallum J C and Prawer S 2011 *New J. Phys.* **13** 045015

[46] Müller T, Aharonovich I, Lombez L, Alaverdyan Y, Vamivakas A N, Castelletto S, Jelezko F, Wrachtrup J, Prawer S and Atatüre M 2011 *New J. Phys.* **13** 075001

[47] Müller T, Aharonovich I, Wang Z, Yuan X, Castelletto S, Prawer S and Atatüre M 2012 *Phys. Rev. B* **86** 195210

[48] Aharonovich I, Castelletto S, Simpson D A, Stacey A, McCallum J, Greentree A D and Prawer S 2009 *Nano Lett.* **9 (9)** 3191−5

[49] Aharonovich I, Castelletto S, Simpson D A, Greentree A D and Prawer S 2010 *Phys. Rev. A* **81** 043813

[50] Castelletto S and Boretti S 2011 *Opt. Lett.* **36 (21)** 4224−26

[51] Deshko Y and Gorokhovsky A A 2010 *J. Low Temp. Phys.* **36 (5)** 465−71

[52] Hsu J H, Su W-D, Yang K-L, Tzeng Y-K and Chang H-C 2011 *Appl. Phys. Lett.* **98** 193116

[53] Lau D W M, Karle T J, Johnson B C, Gibson B C, Tomljenovic-Hanic S, Greentree A D and Prawer S 2013 *APL Materials* **1** 032120

[54] Zaitsev A M 2000 *Phys. Rev. B* **61 (19)** 129−22

[55] Mohammed K, Davies G and Collins A T 1982 *J. Phys. C Condens. Matter* **15** 27−88

[56] Nadolinny V A, Yelisseyev A P, Baker J M, Newton M E, Twitchen D J, Lawson S C, Yuryeva O P and Feigelson B N 1999 *J. Phys. Condens. Matter* **11** 7357−76

[57] Edmonds A M, Newton M E, Martineau P M, Twitchen D J and Williams S D 2008 *Phys. Rev. B* **77** 245205

[58] D'Haenens-Johansson U F S, Edmonds A M, Green B L, Newton M E, Davies G, Martineau P M, Khan R U A and Twitchen D J 2011 *Phys. Rev. B* **84** 245208

[59] Gheeraert E, Kumar A, Bustarret E, Ranno L, Magaud L, Joly Y, Pascarelli S, Ruffoni M, Avasthi D K and Kanda H 2012 *Phys. Rev. B* **86** 054116

[60] Wu E, Jacques V, Zeng H, Grangier P, Treussart F and Roch J-F 2006 *Opt. Express* **14 (3)** 1296−303

[61] Simpson D A, Ampem-Lassen E, Gibson B C, Trpkovski S, Hossain F M, Huntington S T, Greentree A D, Hollenberg L C L and Prawer S 2009 *Appl. Phys. Lett.* **94** 203107

[62] Siyushev P et al. 2009 *New J. Phys.* **11** 113029

[63] *Certain commercial equipment, instruments or materials are identified in this paper to foster understanding; such identification does not imply recommendation or endorsement by the Istituto Nazionale di Ricerca Metrologica, nor does it imply that the materials or equipment are necessarily the best available for the purpose.*

[64] Kitson S C, Jonsson P, Rarity J G and Tapster P R 1998 *Phys. Rev. A* **58** 620−7

[65] Giuntini L, Massi M and Calusi S 2007 *Nucl. Instrum. Methods Phys. Res. A* **576 (2-3)** 266−73

[66] Manfredotti C, Calusi S, Lo Giudice A, Giuntini L, Massi M, Olivero P and Re A 2010 *Diam. Relat. Mater.* **19** 854−60

[67] Popovici G, Wilson R G, Sung T, Prelas M A and Khasawinah S 1995 *J. Appl. Phys.* **77 (10)** 5103−6

[68] Machi I Z, Butler J E, Connell S H, Doyle B P, Maclear R D, Sellschop J P F, Sideras-Haddad E and Rebuli D B 1999 *Diam. Relat. Mater.* **8** 1611−4

[69] Reichart P, Datzmann G, Hauptner A, Hertenberger R, Wild C and Dollinger G 2004 *Science* **306** 1537−40

[70] Gatto Monticone D, Quercioli F, Mercatelli R, Soria S, Borini S, Poli T, Vannoni M, Vittone E and Olivero P 2013 *Phys. Rev. B* **88** 155201

[71] Brouri R, Beveratos A, Poizat J-P and Grangier P 2000 *Opt. Lett.* **25 (17)** 1294−6

[72] Babinec T M, Hausmann B J M, Khan M, Zhang Y, Maze J R, Hemmer P R and Loncar M 2010 *Nat. Nanotechnol.* **5** 195−9

[73] Castelletto S, Degiovanni I P and Rastello M L 2000 *Metrologia* **37** 613−6